\documentclass{epl}

\newcommand{\cats}[1]{\ensuremath{\big|#1\big>}}
\newcommand{\bras}[1]{\ensuremath{\big<#1\big|}}
\newcommand{\nr}[1]{\ensuremath{\big<#1\big>}}

\title{Exact ground states of spin-2 chains.}

\author{M.A.\ Ahrens\inst{1} \and A.\ Schadschneider\inst{1} \and 
J.\ Zittartz\inst{1}}

\institute{
  \inst{1} Institut f{\"u}r Theoretische Physik, Universit\"at zu K\"oln,
50937 K{\"o}ln, Germany
}
                           
\usepackage{amsthm}
\usepackage[reqno,fleqn]{amsmath}
\usepackage{amssymb}
\usepackage{bbm}
\begin{document}

\maketitle

\begin{abstract}
  We use the matrix product approach to construct all optimum ground
  states of general anisotropic spin-2 chains with nearest neighbour
  interactions and common symmetries. These states are exact
  ground states of the model and their properties depend on up to
  three parameters. 
  We find three different antiferromagnetic Haldane phases, one 
  weak antiferromagnetic and one weak ferromagnetic phase. 
  The antiferromagnetic phases can be described as spin liquids with
  exponentially decaying correlation functions. The variety of
  phases found with the matrix product ansatz also gives insight into the
  behaviour of spin chains with arbitrary higher spins. 
  
 \end{abstract}


\section{Introduction}

Haldane's conjecture \cite{Hal.83a,Hal.83b} about the fundamental
difference between integer and half-odd-integer spin chains has
triggered a strong interest in low-dimensional systems with 
arbitrary spin $S$, both theoretically and experimentally.
New materials have been found which can be considered as realizations
of spin chains with $S>\frac{1}{2}$.
For low-dimensional models mean-field theories are usually not
very reliable due to the importance of quantum fluctuations. 
Therefore exact results become very important. They allow to explore
possible phases and study their properties.
Since only few models are solvable exactly, e.g.\ 
by Bethe-Ansatz \cite{Bet.31}, it is necessary to reach out for
alternative methods. Next to numerical
procedures like DMRG \cite{Whi.92,Whi.93} powerful analytical
approaches are available. In this paper we use the idea of
optimum ground states \cite{KSZ.93,BS.95,NZ.96}
which allows to construct systematically {\em exact}
ground states. The method is not restricted to one-dimensional spin 
systems, but can also be generalized to construct ground states for
quantum spin systems in arbitrary dimensions \cite{NKZ.97,NKZ.00}, 
Hubbard models \cite{BS.95} or stochastic processes \cite{DEH.93}.

A prominent example for a spin-2 chain is
(2,2'-bipyridine)tri\-chloro\-manga\-nese(III) where the spin-2 is
carried by $Mn^{3+}$ and the magnetic interaction by the $Mn-Cl-Mn$
orbital overlap. Experiments by Granroth et al.\ \cite{Gra.96} 
show antiferromagnetic behaviour with exponentially
decaying correlation functions. In the present letter we will construct 
states which exhibit exactly this behaviour with a non-degenerate
ground state. Not only antiferromagnetic phases can be realized for a 
spin-2 chain but also weak antiferromagnetic and weak ferromagnetic ones.
As the method of optimum ground
states allows us to calculate ground state expectation values and
properties for these states, it provides a good overview over the large
variety of phases encountered in spin-2 chains and their properties.

In the following we consider quantum spin chains described by a
translational invariant Hamiltonian $H=\sum_{\langle i,j\rangle}h_{ij}$
with nearest neighbour interactions $h_{ij}$.
   
{\bf Definition.} A {\em global} ground state $\cats{\Psi_0}$ is called
\emph{optimum ground state} (OGS) of $H=\sum_{\langle i,j\rangle}h_{ij}$, 
if the global ground state energy $E_0$ is just the sum of the local
ground state energy values $\epsilon_0$ of the local Hamiltonians
$h_{ij}$. 

In other words: If the local ground state conditions
$h_{ij}\cats{\Psi_0}=0$ are complied (which can be achieved by
adding the constant $-\epsilon_0$ to the local interaction $h_{ij}$) 
this yields the equivalence
\begin{equation}\label{e.ogs}
H\cats{\Psi_0}=0\quad  \Longleftrightarrow \quad 
h_{ij}\cats{\Psi_0}=0,  \qquad \forall{i,j}
\end{equation}
since then $0$ is a lower bound for $E_0$.
The r.h.s.\ of (\ref{e.ogs}) can be used as local conditions for the
existence of an optimum ground state and its realization in terms of a
\emph{matrix product ground state} (MPG) (see below). Obviously OGS
are states without finite-size corrections. In \cite{KSZ.93,KSZ.91,KSZ.92}
the matrix product ground state approach has been described in more detail. 

For the spin-1 chain \cite{KSZ.93} it is possible to construct one type 
of matrix product ground state for a Hamiltonian with the
three symmetries 1) rotational invariance in the ($x,y$)-plane, 2)
spin flip invariance, 3) translation and parity invariance.
In the case of spin-2 chains with these symmetries and nearest
neighbour interaction we find five 
different non-trivial MPGs. It can be understood easily that there
exist no more than these five MPG except for exponentially degenerate 
ones \cite{upcoming}.

Usually it is most common to find exact or
approximate ground states or even the full spectrum for a {\em given}
Hamiltonian. The modus operandi we use here is the other way around. First
we construct the optimum ground states for a given system and then we 
determine the subspace of Hamiltonians for which these
are the ground states. Describing the local Hamiltonians $h_{ij}$ in
terms of local eigenstates $\cats{v_k}$ of
$\hat{S}^z_i+\hat{S}^z_j$ (eigenvalue $s$)
and the parity operator $\hat{P}_{ij}$ (eigenvalue $p$) and the spin 
flip invariance lead to $h_{ij}=\sum_{k} \lambda_k\cats{v_k}\bras{v_k}$ 
(see~\cite{Ahr.01}).
For some sets of quantum numbers $s,p$ the $\cats{v_k}$ are not determined 
completely by the imposed symmetries. Additional superposition parameters 
control the "orientation" of the basis states $\cats{v_k}$ in their 
respective subspace.
Requiring that a given OGS is an exact ground state of $H$ leads to
restrictions on the $\lambda_k$ and the superposition parameters 
of the $\cats{v_k}$. 
For a spin-2 system with nearest neighbour
interaction and above mentioned symmetries the total number of parameters 
is 22 \cite{NKZ.00}. Two parameters, the energy off-set and a scale, 
are trivial which leads effectively to a 20 parameter model. 
For the most general model there are 
25 local energy eigenstates.

As expected such a model with 20 parameters in
general has very complicated ground states of the global Hamiltonian $H$.
The concept of optimum ground states and its realization through MPG
yields a class of structural simple states.
For these it is possible to calculate ground state expectation
values for arbitrary operators.
The isotropic valence-bond-solid (VBS) model  \cite{AKLT.88} emerges as a 
special case and the properties of this model turn out to be
generic for a more general subspace of the 20 parameter space.

 
\section{The matrix product ground state}

Following the procedure explained in \cite{KSZ.93} one can define
several matrices which are suited for constructing matrix product
ground states of the spin-2 chain. Such a matrix consist of single-spin 
states at site $i$ as their elements and the product of two such 
matrices is defined as a matrix product with the tensor product
$\otimes$ for the spin states
\begin{equation}
(m^{(i)}\cdot m^{(i+1)})_{\mu \nu}=\sum_{k} m^{(i)}_{\mu k} 
\otimes m^{(i+1)}_{k\nu}.
\end{equation}
{\bf Definition.} A \emph{matrix product ground state} (MPG) is a
global ground state of a spin chain of length $L$, which for 
periodic boundary conditions can be written in the form
\begin{equation}
\cats{\Psi_0}=\operatorname{tr}\left( m^{(1)}\cdot m^{(2)}\cdot m^{(3)}
\cdots m^{(L)}\right), 
\end{equation}
where $\operatorname{tr}$ stands for  ``trace over the matrix space", 
~\emph{i.e.} $\sum_{\mu}\left(m^{(1)}\cdot m^{(2)}\cdots m^{(L)}
\right)_{\mu\mu}$.
For a MPG to satisfy the above optimum ground state condition,
$h_{i,i+1}\cats{\Psi_0}=0$ must hold for all $i$. 
Due to the product structure of the ground state $\cats{\Psi_0}$ this
reduces to
\begin{equation}\label{e.gsc}
h_{i,i+1}\left(m^{(i)} \cdot m^{(i+1)}\right)_{\mu \nu}=0 
\qquad\text{ for all } i,\mu,\nu. 
\end{equation}
 
This implies that each element of the product matrix 
$\left(m^{(i)} \cdot  m^{(i+1)}\right)$ 
is a local ground state (of $h_{i,i+1}$). These
conditions lead to several restrictions on the parameters from the
most general model. So the matrix product ground state becomes a
global ground state of the specified parameter subspace. 
The MPG itself turns out to depend on up to three parameters. 
To calculate expectation values and other physical properties
a transfer matrix approach ~\cite{KSZ.92} can be used.
Results thus obtained are presented in the following sections.


\section{MPGs on the spin-2 chain}

We have investigated systematically matrix product states on a spin-2 
chain with unique or finitely degenerate ground states and periodic
boundary conditions. We found five MPGs with non-trivial product
structure and different properties. 
In the following only the most significant properties will be presented. 
A more complete account will be published elsewhere ~\cite{upcoming}.

\section{Haldane-Antiferromagnet-A} 

\begin{figure}
\twofigures[width=7cm]{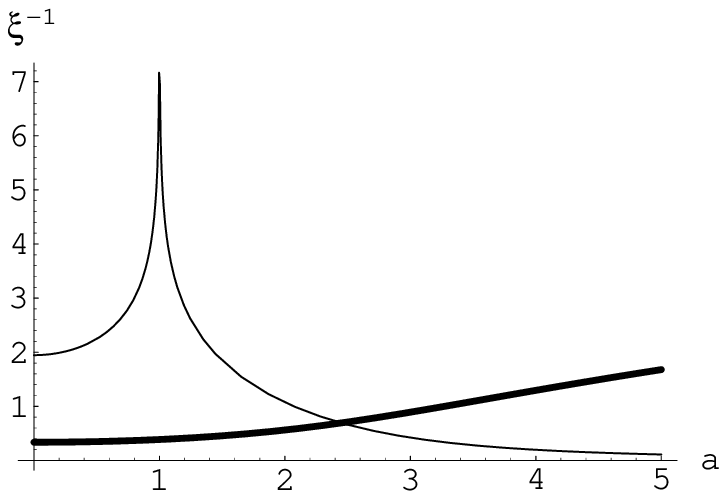}{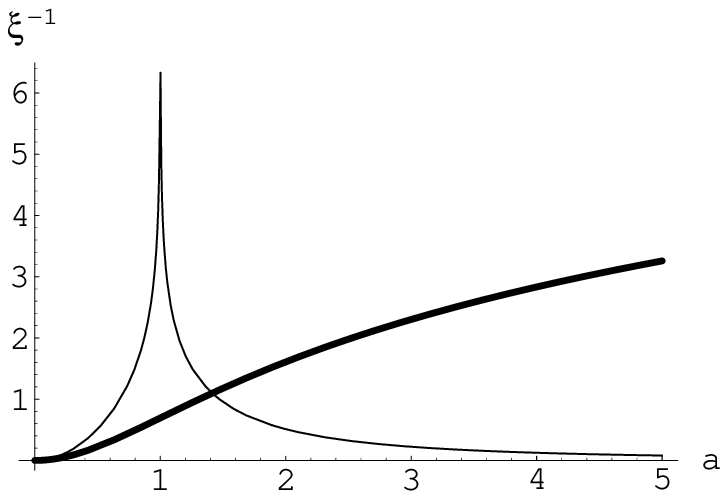}
\caption{Longitudinal (thin) and transversal correlation 
length $\xi$ for $x=-3$ and $\gamma=-2$. This special choice of parameter 
values 
includes the isotropic point for $a=\sqrt{6}$ where the correlation lengths
are equal.}\label{f.2}
\caption{For $x=0$, $\gamma=1$ the correlation lengths diverge
in the limit $a\to 0$. Note that the crossing point does not correspond
to an isotropic Hamiltonian.}\label{f.1}
\end{figure}

The first MPG is defined by the homogeneous product 
$\cats{\Psi_0}(a,x,\gamma)=\operatorname{tr}
\left(\prod_{i}^{L}m^{(i)}\right)$ of matrices 
\begin{equation}
\label{eq_HAF1a}
  m^{(i)}= \left(\begin{tabular}{rrr} $\cats{0}_i$ &
      $\sqrt{x}\cats{1}_i$ & $a\cats{2}_i$ \\ 
      $\sqrt{x}\cats{\overline{1}}_i$ & $\gamma \cats{0}_i$ &
      $\sqrt{x}\cats{1}_i$\\ $a \cats{\overline{2}}_i$ & $\sqrt{x}
      \cats{\overline{1}}_i$ & $ \cats{0}_i$
\end{tabular}\right).
\end{equation} 
This representation with three continuous, real parameters $a,x,\gamma$
uses the canonical spin-2 basis states, ~\emph{i.e.}\ the eigenstates of the 
$\hat{S}^z$-operator $\hat{S}^z_i\cats{s^z}_i=s^z\cats{s^z}_i$ and
$\hat{S}^z_i\cats{\overline{s^z}}_i=-s^z\cats{\overline{s^z}}_i$ at
site $i$. 
Since all $h_{i,i+1}$ should be the same (for all $i$) we have to 
solve (\ref{e.gsc}) for one arbitrary $i$ only. These conditions lead to 
a 12 parameter subspace and an additional trivial parameter for the scale.
In general inequalities must hold for these parameters. For more details 
see \cite{upcoming}. 
This parameter space includes the isotropic point. 
The ground state itself
depends on the three parameters $a,x,\gamma$ and is \emph{unique},
which can be proven rigorously by complete induction \cite{Ahr.01}.  
For the parameter set $x=-3, a=\sqrt{6}, \gamma=-2$ (see fig.~\ref{f.2}) 
one obtains the 
isotropic point of the model where the matrix product ground state 
reproduces the VBS state \cite{AKLT.88}. The 
corresponding Hamiltonian can be written as sum of the projection 
operators onto the  $S^z=3$ and $S^z=4$ multiplets.
For any other set of 
parameters the model has an anisotropy along the $z$-axis. However, in
any case the ground state is unique as long as $a\neq 0$. For
$a=0$ the matrix product Ansatz (MPA) is not unique; the degeneracy
grows with system size as $3^L$. This case is not considered in the
following. 

The ground state defined by (\ref{eq_HAF1a}) is antiferromagnetic 
in the sense that all single-site magnetisations vanish:
$\nr{\hat{S}^z}\equiv\nr{\hat{S}^x}\equiv\nr{\hat{S}^y}\equiv0$. 
It follows that the square of the fluctuation is simply given by
\begin{equation}
\left(\Delta S^z \right)^2=\nr{(\hat{S}^z)^2}-\nr{\hat{S}^z}^2=
\nr{(\hat{S}^z)^2}\in [0,4].
\end{equation}
The maximum $\left(\Delta S^z \right)^2=4$ is reached in the limit  
$|a|\rightarrow \infty$ and $\Delta S^z=0$ for $|\gamma|\rightarrow \infty$.
In the latter case the dominant contribution is proportional to
$\gamma^L\cats{000\cdots0}$. The 2-site correlation functions decay
exponentially to zero. The longitudinal correlation function is given by
\begin{equation}
\nr{\hat{S}_1^z\hat{S}^z_r}=\nr{\hat{S}_1^z\hat{S}^z_2}
\left(\operatorname{sign}\left(1-a^2\right)\right)^r
\operatorname{e}^{-(r-2)/\xi_l},\qquad (r\geq 2),
\end{equation}
with the longitudinal correlation
\begin{equation}
\xi_l^{-1}=\ln\left|\frac{\lambda}{1-a^2}\right|, \qquad 
\lambda=\frac{1}{2}\left((1+a^2+\gamma^2)+\sqrt{(1+a^2-\gamma)^2+8x^2}\right).
\end{equation}
The first part of the function is the expectation value of the nearest
neighbour correlation which is antiferromagnetic
($\nr{\hat{S}_1^z\hat{S}^z_2}\le 0$) and varies from $0$ (for $\gamma
\rightarrow \infty$) to $-4$ (for $a^2 \rightarrow \infty$) \cite{Ahr.01}. 
For large $a$ the correlation function alternates in $r$ as an easy axis
anisotropy is observed. The transversal correlation function decays
exponentially, too,
\begin{equation}
\nr{\hat{S}^x_1\hat{S}^x_r}= \nr{\hat{S}_1^x\hat{S}^x_r}\left(|x|
+\gamma\right)^r\operatorname{e}^{-(r-2)/\xi_t}, \qquad \xi_t^{-1}
=\ln\left|\frac{\lambda}{|x|+\gamma}\right|.
\end{equation}

Both correlation lengths $\xi_l$ and $\xi_t$ are finite (except for $a=x=0$,  
see fig.~\ref{f.1}) and so the model is non-critical. 
The $a$-dependence (see fig.~\ref{f.2} \&~\ref{f.1}) 
looks quite similar to the spin-1 MPG correlation lengths presented in
\cite{KSZ.93}. For large $a$ the anisotropy is of easy-axis type 
with $\xi_l>\xi_t$ whereas for small $a$ it is of easy-plane type.
Indeed this MPG is the analogue of the matrix product ground state of 
the spin-1 chain, e.g.\ the isotropic point (see fig.~\ref{f.2}) is 
included in this phase for both spin chains. For both chains the 
antiferromagnetic phases (including the two antiferromagnetic ones 
described in the next subsections) are Haldane phases with
1) a unique ground state, 2) exponentially decaying 
correlation functions, and 3) a gap to the first excited state.
Here we only have shown 1) and 2) of this scenario, but the existence of a
gap can be shown along the lines of the proof outlined in \cite{AKLT.88}. 


\section{Haldane-Antiferromagnet-B}

Different from the spin-1 chain it is possible to construct further
matrix product ground states for the spin-2 case. A simple one is a
homogeneous product of matrices
\begin{equation}
\label{eq_HAF1b}
m=
\left(\begin{tabular}{rr}
$\cats{0}$ & $\sqrt{a}\cats{1}$  \\
$\sqrt{a}\cats{\overline{1}}$ & $\sigma\cats{0}$  
\end{tabular}\right),\qquad a\in\mathbb{R},\sigma\pm1,
\end{equation}
which in contrast to the Haldane-antiferromagnet-A depends
on one continuous parameter $a$ and one discrete parameter $\sigma$.
The state $\cats{\Psi_0}(a)=\operatorname{tr}\left(\prod_{i}^L
  m_i\right)$ looks quite similar to the one presented in
\cite{KSZ.93} for the spin-1 case.
Again it can be shown that it is a unique ground state of $H$
in the appropriate subspace.
Even the expectation values of all combinations of the $\hat{S}^z_i$ 
operators look the same as in the spin-1 case, 
only expectation values of $\hat{S}^+_i$ or
$\hat{S}^-_i$ operators vary by a factor. The properties of this
phase are much the same of the one before and so we report some
expectation values only. The single-site magnetisations vanish and the
square of the fluctuation just as the longitudinal correlations
function are the same as the ones for the spin-1 chain \cite{KSZ.93}.
Different from the spin-1 chain the transversal correlation function
reads:
\begin{equation}
\nr{\hat{S}^x_1\hat{S}^x_r}=-3|a|(\operatorname{sign}(a)-\sigma)\cdot
\left(\frac{\sigma}{1+|a|}\right)^r, \qquad 
\xi_t^{-1}=\operatorname{ln}\left(1+|a|\right).
\label{xkorr_HAF1b}
\end{equation}
The correlation length $\xi_t$ is the same as in \cite{KSZ.93},
only with a different prefactor.


\section{Haldane-Antiferromagnet-C}

Replacing in (\ref{eq_HAF1b}) the spin states $\cats{1}$ by 
$\cats{2}$ and $\cats{\overline{1}}$ by $\cats{\overline{2}}$ 
we obtain a third matrix product ground state,
which is the \emph{unique} ground state in a 16 parameter subspace of
the most general 20 parameter model. The expectation values of
$\hat{S}^z$-operators are the same as the ones above, except of a
factor coming from the higher $\hat{S}^z$-eigenstates ($\cats{\pm2}$
instead of $\cats{\pm1}$). For other operators, e.g.\ 
$\hat{S^x}=\frac{1}{2}(\hat{S}^++\hat{S}^-)$,  the method of
transfer matrices shows that elements like
$\bras{m_{\mu,\nu}}\hat{S}^+\cats{m_{\tilde{\mu},\tilde{\nu}}}$ become
relevant. Because the matrix contains elements proportional to
$\cats{\overline{2}}, \cats{0}$ and $\cats{2}$ only, this always leads
to zero and therefore $\nr{\hat{S}^x_1\hat{S}^x_r}\equiv 0$. However,
the expectation values of biquadratic operators of this type do not
necessarily vanish, e.g.
\begin{equation}
\nr{(\hat{S}^x_1\hat{S}^x_r)^2}=\left(\frac{3+|a|}{1+|a|}\right)^2+3(
\operatorname{sign}(a)+\sigma)|a|\left(\frac{\sigma}{1+|a|}\right)^r.
\end{equation}


\section{Weak-Antiferromagnet}

In the following we will show that also other antiferromagnetic
phases can be realized by the matrix product technique, e.g.\ a
phase characterized by a vanishing total magnetisation of two neighbouring 
sites $\langle S_j^z+S_{j+1}^z\rangle =0$, but with finite single-site 
magnetisation $\langle S_j^z\rangle \neq 0$. If the sublattice is not
fully polarized, ~\emph{i.e.}\ $\left|\langle S_j^z\rangle\right| < 2$,
the corresponding state will be denoted as \emph{weak antiferromagnet}.

In order to realize a weak antiferromagnet using the MPA we introduce
the two matrices
\begin{equation}
m=
\left(\begin{tabular}{rrr}
$\cats{1}$ & $x\cdot\sqrt{a}\cats{2}$ \\
$\sqrt{a} \cats{0}$ & $\cats{1}$
\end{tabular}\right),\quad
g=
\left(\begin{tabular}{rrr}
$\cats{\overline{1}}$ & $\sqrt{a}\cats{0}$ \\
$x \cdot \sqrt{a} \cats{\overline{2}}$ & $\cats{\overline{1}}$
\end{tabular}\right).
\end{equation}
Using these matrices, two different MPGs can be constructed by assigning
the matrices $g$ and $m$ to different sublattices:
$\cats{\Psi_0^{(1)}}=\operatorname{tr}\left(\prod_{i}^{L/2}m_{2i-1}\cdot
  g_{2i}\right)$ and
$\cats{\Psi_0^{(2)}}=\operatorname{tr}\left(\prod_{i}^{L/2}g_{2i-1}\cdot
  m_{2i}\right)$. In the first case to every odd lattice site a matrix 
$m$ is attached and a matrix $g$ to the even numbered sites. In the 
second case the situation is reversed. 
 Therefore the ground state is twofold degenerate and
each ground state depends on the parameter $a$.
The single-site magnetisation alternates from lattice site to lattice
site and can be written as
\begin{equation}
\nr{S^z_m}=-\nr{S^z_g}=1+\frac{|a|(x^2-1)}{\sqrt{4+a^2(x^2-1)^2}}.
\end{equation}
For $|x|=1$ the
sub-lattice magnetisation becomes $\nr{S^z_m}=-\nr{S^z_g}=1$. In the
limit $|x\cdot a|\rightarrow \infty$ the magnetisation is
$\nr{S^z_m}=0$ for values $|x|<1$ and $\nr{S^z_m}=2$ for $|x|>1$. For
$|x\sqrt{a}|\rightarrow \infty$ but $|a|\rightarrow 0$ a strict N{\'e}el 
order with $\cats{\Psi_0^{(1)}}=\cats{2\overline{2}2\ldots \overline{2}}$ and 
$\cats{{\Psi}_0^{(2)}}=\cats{\overline{2}2\overline{2}\ldots 2}$
is realized. 
The magnetisations in $x$- and $y$-direction vanish for the whole
subspace
\begin{equation}
\nr{S^x_m}=\nr{S^y_m}=\nr{S^x_g}=\nr{S^y_g}=0.
\end{equation}
The fluctuation $\Delta S^z$ varies in the range of 0 to 1. It takes
its maximum for $|x|=1$ and $|a|\rightarrow \infty$.
The longitudinal correlation function is given by
\begin{equation}
\nr{S_1^zS_r^z}=(-1)^{r+1}\left(\nr{S^z_m}^2+A_l
\operatorname{e}^{-r/\xi_l}\right)\qquad (r\geq 2)
\end{equation}
and alternates in $r$. It decays exponentially to a constant 
contribution $\nr{S^z_m}^2$ which is a result of translation 
invariance breaking. The amplitude $A_l$ is positive and depends
on the parameters $a$ and $x$ \cite{Ahr.01}.

The transversal correlation function decays exponentially to zero. 
Again, the correlation lengths stay
finite for all values. In the limits $a\to 0+$ and $a\to 0-$ 
both correlation lengths, the transversal and the longitudinal, diverge. 
For $a=0$ the ground state degeneracy grows exponentially with the chain 
length. 


\section{Weak-Ferromagnet}

The next two matrix product ground states look quite similar to the
ones above. The main difference is that the translational invariance 
is not broken here. With the matrices
\begin{equation}
m=
\left(\begin{tabular}{rrr}
$\cats{1}$ & $\sqrt{a}\cats{2}$ \\
$\sqrt{a} \cats{0}$ & $\sigma \cats{1}$
\end{tabular}\right),\quad
g=
\left(\begin{tabular}{rrr}
$\cats{\overline{1}}$ & $\sqrt{a}\cats{0}$ \\
$\sqrt{a} \cats{\overline{2}}$ & $\sigma \cats{\overline{1}}$
\end{tabular}\right).
\end{equation}
the two MPG
$\cats{\Psi_0^{(m)}}=\operatorname{tr}\left(\prod_{i}^L m_i\right)$ and
$\cats{\Psi_0^{(g)}}=\operatorname{tr}\left(\prod_{i}^Lg_i\right)$
can be constructed. 
Due to the spin flip symmetry 2) one has to ensure that both states are
ground states of $H$ which then is twofold degenerate.
Just as expected the single-site magnetisation in $z$-direction reads
$\nr{\hat{S}^z}=-\nr{\hat{S}^z}=1$ and in $x$- resp.\ in $y$-direction
$\nr{\hat{S}^x}=\nr{\hat{S}^y}=0$.  The
square of the fluctuation $\Delta \hat{S}^z$ shows the same
$a$-dependence as the one from the spin-1 MPG
\begin{equation}
\left(\Delta \hat{S}^z \right)^2=\frac{|a|}{1+|a|}, \qquad (r\geq 2).
\end{equation}
The correlation lengths are the same as in 
the spin-1 case, 
\begin{eqnarray}
\xi_l^{-1}=\operatorname{ln}\left|\frac{1+|a|}{1-|a|}\right|,\\
\xi_t^{-1}=\operatorname{ln}\left(1+|a|\right).
\end{eqnarray} 
Differences can be found in the longitudinal correlation function
itself. The possibility to choose either $\cats{\Psi_0^{(m)}}$ or
$\cats{\Psi_0^{(g)}}$ leads to a finite contribution to the correlation
function
\begin{equation}
\nr{\hat{S}^z_1\hat{S}^z_r}=1-\frac{a^2}{(1-|a|)^2}
\left(\frac{1-|a|}{1+|a|}\right)^r.
\label{corr_WF}
\end{equation}
This partially polarized state is structurally similar to an 
antiferromagnetic state. It can be viewed as a Haldane-type state,
but with finite magnetization. This is also reflected in the 
behaviour of the correlation function (\ref{corr_WF}) which
agrees with that of the Haldane-AF-B up to the long-range order
part which is zero in the latter case.


\section{Conclusion}

The results presented here show that for a spin-2 chain the spectrum
of possible MPG is much larger than for a spin-1 chain. 
We found three antiferromagnetic phases with unique ground state,
exponentially decaying correlation functions and finite excitation gap.
Therefore the corresponding phases can be classified as Haldane phases
or spin liquids. Their structure is similar to that found in the
case of the spin-1 chain \cite{KSZ.93}. 
Similar antiferromagnetic phases can not be constructed for the
spin-$\frac{3}{2}$ chain \cite{NZ.96}. However, in this case a weak 
antiferromagnetic and a weak ferromagnetic phase exist. 
Related states can also be constructed in the present spin-2 case.
The weak antiferromagnet shows exponentially decaying correlation
functions with long-range order. The ground state is twofold
degenerate reflecting the breaking of translation invariance and
leading to a finite sublattice magnetisation.
The weak ferromagnet has a twofold degenerate ground state,
but full translation invariance. The
magnetisation per site in $z$-direction takes the constant value 1,
 which is just half of the fully polarised
ferromagnet. Interestingly, despite the finite magnetisation,
this state is structurally similar to an antiferromagnetic state.

We believe that the exact results presented here are generic for
spin-2 chains in a similar way as previous results \cite{KSZ.93,AKLT.88} 
are for spin-1 chains. Here for the isotropic case, ~\emph{i.e.}\ the
bilinear-biquadratic chain, an exact solution using OGS is only possible
for one point, the AKLT-chain \cite{AKLT.88}. However, a whole
extended phase exists with the same properties. In this sense the
exact results allow to study the generic properties of such phases
without having to rely on approximate or numerical methods.

Finally we like to point out that the results found here can be extended
to even larger spins just as the Haldane-antiferromagnets-B and C can
be viewed as analogues of the spin-1 states.


\end{document}